\documentclass[]{aa}
\graphicspath{{../}{figs/}}

\usepackage{CJK}
\usepackage{tabularx}
\usepackage{booktabs}
\usepackage{inputenc}
\usepackage{textcomp}
\usepackage{subfigure}
\usepackage{amsmath}
\usepackage{threeparttable}
\defcitealias{Dou25}{D25}

\begin{document}
\begin{CJK*}{UTF8}{gbsn}

\title{The dual effect of group-scale environments on galaxy quenching during cluster infall: pre-processing and protection}

\author{
Haoran Dou (窦浩然)\inst{\ref{inst1}}
\and Heng Yu (余恒)\inst{\ref{inst1}}\thanks{Corresponding author. E-mail: yuheng@bnu.edu.cn}
\and Xiaolan Hou (侯晓岚)\inst{\ref{inst1}}
}

\institute{School of Physics and Astronomy, Beijing Normal University Beijing 100875, PR China \label{inst1}}

\abstract
{It is well established that the cluster environment effectively quenches star formation in member galaxies.}
{We aim to explore how the accretion path of infalling galaxies influences the cluster-driven quenching process.}
{We compiled a large spectroscopic galaxy sample around 25 low-redshift, X-ray luminous massive clusters.
We identified cluster substructures using the Blooming Tree algorithm and distinguished between galaxies accreted as part of group-scale structures and those accreted in isolation.
The infall process was quantified using an infall proxy, $d_{\rm R}$, defined in the $R$--$V$ diagram.}
{Along the infall process, the quiescent fraction remains approximately constant at the outskirts and then increases steadily toward cluster center, with a transition occurring around $d_{\rm R}\sim 2.5$.
We find that group-associated galaxies follow a distinct quenching track compared to isolated galaxies, indicating a dual effect of group-scale environments. 
At the early infall stages, group galaxies exhibit a higher quiescent fraction, consistent with ``pre-processing'' in group-scale halos. 
However, after entering the cluster environment, the rise in their quiescent fraction is delayed to smaller $d_{\rm R}$ compared to isolated galaxies.
This suggests a phenomenological ``protection'' effect, in which group-scale halos buffer member galaxies against rapid cluster-driven quenching.}
{We conclude that group-scale environments affect quenching in two ways: via pre-processing prior to cluster infall, and through a subsequent protection effect within the cluster environment. }

\titlerunning{The Dual Effect of Group Environments}
\authorrunning{Dou et al.}
\maketitle
\nolinenumbers

\section{Introduction}
\label{sec:intro}

Galaxies in the Universe exhibit clear bimodal distributions in many physical properties, such as photometric colors, star formation rate (SFR), and morphology \citep[e.g.,][]{Strateva_2001, Bell_2004, Peng_2010, Schawinski_2014}.
A central goal in galaxy evolution studies is to understand how star-forming galaxies are quenched to the quiescent population, and how this transition depends on both galaxy properties and the environment.

It is well known that quenching is driven by a combination of internal processes and external environmental effects \citep{Peng_2010, Cortese_2021, DeLucia2025}.
Internal processes, primarily stellar and active galactic nuclei (AGN) feedbacks, can heat or expel cold gas and subsequently suppress star formation \citep[e.g.,][]{Cicone_2014, Wang23, Belli_2024, Wang25}. These processes are closely related to the stellar and black hole masses of galaxies \citep{Croton_2006, Terrazas_2020, 2024Wang}, producing more quiescent galaxies at the high-mass end.
In addition, red and quiescent galaxy populations are more frequently found in dense environments, such as groups and clusters \citep{Dressler_1980}, where hydrodynamical interactions with the intracluster medium (ICM) and gravitational interactions with other galaxies together regulate the gas supply \citep{van_den_Bosch_2008, 2010vonderLinden, 2012Smith, Boselli_2014, Wright_2022}.  
In less dense regions, such as groups and filaments, moderate gas stripping removes only the circumgalactic medium (CGM) of galaxies, cutting off the replenishment of the cold gas reservoir. This process, known as ``starvation'' (or ``strangulation''), ultimately leads to quenching over several gigayears \citep[e.g.,][]{Larson_1980_starvation, Balogh_2000, Kawata_2008, vandeVoort_2017, Trussler_2020}.
In denser environments, such as the core regions of clusters, severe ram pressure stripping can directly remove the interstellar medium (ISM), resulting in very rapid quenching on timescales of $\sim0.1-1$ Gyr \citep{Gunn_1972_RPS, Yun_2019, Roberts_2019, Boselli_2022, Rohr_2023}.
Galaxy-galaxy interactions, including tidal stripping \citep{Bournaud_2004, Koopmann_2004, Wang_2022}, harassment \citep{Moore_1996, Moore_1998}, and mergers \citep{Spitzer_1951, Teyssier_2010}, can also strongly contribute to quenching.

Galaxy quenching in clusters is particularly complex because various mechanisms often operate simultaneously \citep{DeLucia2025}. Moreover, the diverse accretion histories of galaxies further complicate the quenching picture \citep{2009Berrier, 2009McGee, 2012Smith}.
According to the standard cold dark matter (CDM) paradigm, large-scale cosmic structures form hierarchically, meaning galaxies may traverse multiple environments before reaching present-day clusters \citep{Zeldovich_1970, 1984Blumenthal, 1985Davis, Colberg2005, Libeskind2018, Primack24}.
Some galaxies are initially accreted into lower-density structures (such as sheets, filaments, and groups), where ``pre-processing'' can already alter their gas content and star formation activity \citep[e.g.,][]{Fujita_2004, 2009McGee, Haines_2013_slowquench, 2013Bahe, Kleiner_2021, Piraino_2024, Lopes24}. These galaxies may subsequently enter the cluster as part of a bound system, potentially appearing as substructures in the cluster outskirts \citep{ Yu2015, Balestra2016, Yu2016, Yu2018}.
In contrast, other galaxies are accreted individually from the field and may remain relatively unaffected until they are directly exposed to the cluster environment \citep{2009Berrier}. 
Despite extensive work on average quenching trends in clusters, the dependence of quenching on the accretion path (e.g., group versus isolated infall) remains less constrained observationally.
Furthermore, it is unclear whether group-scale environments continue to influence galaxies after they enter the clusters.

Tracing the infall stage of observed galaxies offers a natural route to address this question. 
While simulations provide full 6D kinematics and orbital histories of galaxies \citep[e.g.,][]{Oman_2013}, observations are limited to projected clustercentric distances and line-of-sight (LOS) peculiar velocities. 
This observable kinematic parameter space is called ``$R$--$V$ diagram''\footnote{
While this parameter space is frequently referred to as the Projected Phase Space (PPS) in numerical simulation studies, we retain the term ``$R$--$V$ diagram'' in this work, following the historical convention in observational astronomy. Geometrically, the diagram is constructed by selecting and combining specific coordinates from the 6D phase space, rather than being a mathematical projection that involves integration over the remaining dimensions. From the perspective of classical mechanics, $R$ and $V$ are not canonically conjugate pairs, making the term ``phase space'' theoretically inaccurate.}
, where $R$ represents the normalized projected clustercentric radius and $V$ denotes the normalized LOS velocity relative to the cluster center. The $R$--$V$ diagram has been widely used to statistically infer infall times based on calibration from simulations \citep[e.g.][]{Oman_2013, Oman_2016, Rhee_2017, Pasquali_2019, Dou25}.
Such approaches have revealed systematic evolutionary trends along the infall path for observed cluster galaxies \citep[e.g.][]{Noble_2013, Noble_2016, Sampaio_2021, Qu_2023, Kim_2023, Brambila_2023, Oxland_2024, Sampaio_2024}.
In this paper, we combine the infall process quantified in the $R$--$V$ diagram with substructure identification to investigate how group-scale environments modulate galaxy quenching during cluster infall.

This paper is organized as follows. Section~\ref{sec:data} presents the data sets used in this work. Section~\ref{sec:methods} details the methods used to quantify the infall stage in the $R$--$V$ diagram and to identify substructures within the clusters. Section~\ref{sec:results} reports the results on galaxy quenching across different environments.
We discuss the implications in Section~\ref{sec:discussion} and summarize our conclusions in Section~\ref{sec:summary}.
Throughout this paper, we use the term ``quiescent'' to describe the star formation state of galaxies and ``quenching'' to denote the evolutionary process, although the two concepts are often used interchangeably in the literature.

\section{Data} 
\label{sec:data}

\begin{table*}
    \centering
    \caption{Basic information of the clusters sample.}
    \label{tab:cls_cat}
\begin{threeparttable}
\begin{tabularx}{0.9\textwidth}{lccccccccrrr}
\toprule
 & MCXC ID & Abell ID & R.A. & Decl. & z & $R_{500}$ & $M_{500}$ & $\rm N_{tot}$\tnote{a} & $\rm N_{data}$\tnote{b} & $\rm N_{core}$\tnote{c} \\
 &   &   & [deg] & [deg] &   & [Mpc] & [$10^{14}\rm M_\odot$] &   &   \\
\midrule
1 & J1602.3$+$1601 & A2147 & 240.578 & 16.020 & 0.0353 & 0.935 & 2.4 & 1551 & 1276 & 249 \\
2 & J1558.3$+$2713 & A2142 & 239.586 & 27.227 & 0.0894 & 1.380 & 8.1 & 1230 & 1194 & 170 \\
3 & J0041.8$-$0918 & A0085 & 10.459 & -9.302 & 0.0555 & 1.210 & 5.3 & 500 & 487 & 134 \\
4 & J1522.4$+$2742 & A2065 & 230.610 & 27.709 & 0.0723 & 1.048 & 3.5 & 812 & 727 & 126 \\
5 & J1217.6$+$0339 & - & 184.419 & 3.663 & 0.0766 & 1.055 & 3.6 & 429 & 414 & 106 \\
6 & J1521.2$+$3038 & A2061 & 230.321 & 30.640 & 0.0777 & 0.977 & 2.9 & 505 & 315 & 106 \\
7 & J1348.8$+$2635 & A1795 & 207.221 & 26.596 & 0.0622 & 1.224 & 5.5 & 975 & 893 & 105 \\
8 & J1523.0$+$0836 & A2063 & 230.773 & 8.602 & 0.0355 & 0.902 & 2.2 & 877 & 815 & 96 \\
9 & J1516.7$+$0701 & A2052 & 229.183 & 7.019 & 0.0353 & 0.947 & 2.5 & 1304 & 1214 & 96 \\
10 & J0056.3$-$0112 & A0119 & 14.076 & -1.217 & 0.0442 & 0.941 & 2.5 & 740 & 511 & 91 \\
11 & J1336.1$+$5912 & A1767 & 204.025 & 59.208 & 0.0701 & 0.919 & 2.4 & 421 & 356 & 90 \\
12 & J2354.2$-$1024 & A2670 & 358.556 & -10.413 & 0.0765 & 0.911 & 2.3 & 324 & 301 & 85 \\
13 & J1341.8$+$2622 & A1775 & 205.474 & 26.372 & 0.0724 & 0.933 & 2.5 & 364 & 337 & 76 \\
14 & J1521.8$+$0742 & - & 230.458 & 7.709 & 0.0442 & 0.947 & 2.5 & 869 & 814 & 74 \\
15 & J1709.8$+$3426 & A2249 & 257.453 & 34.441 & 0.0802 & 0.964 & 2.8 & 431 & 376 & 72 \\
16 & J1353.0$+$0509 & A1809 & 208.254 & 5.155 & 0.0788 & 0.866 & 2.0 & 282 & 270 & 68 \\
17 & J1330.8$-$0152 & A1750 & 202.708 & -1.873 & 0.0852 & 0.997 & 3.1 & 394 & 366 & 65 \\
18 & J1712.7$+$6403 & A2255 & 258.197 & 64.061 & 0.0809 & 1.068 & 3.7 & 469 & 172 & 64 \\
19 & J1132.8$+$1428 & A1307 & 173.221 & 14.469 & 0.0834 & 0.993 & 3.0 & 382 & 357 & 59 \\
20 & J1111.6$+$4050 & A1190 & 167.910 & 40.843 & 0.0794 & 0.869 & 2.0 & 532 & 504 & 57 \\
21 & J1113.3$+$0231 & A1205 & 168.336 & 2.532 & 0.0780 & 0.874 & 2.0 & 336 & 314 & 52 \\
22 & J1349.3$+$2806 & A1800 & 207.340 & 28.104 & 0.0748 & 0.903 & 2.3 & 403 & 345 & 48 \\
23 & J1620.5$+$2953 & A2175 & 245.132 & 29.895 & 0.0972 & 0.938 & 2.6 & 215 & 204 & 45 \\
24 & J1702.7$+$3403 & A2244 & 255.679 & 34.062 & 0.0953 & 1.129 & 4.5 & 565 & 360 & 39 \\
25 & J1359.2$+$2758 & A1831 & 209.823 & 27.973 & 0.0612 & 0.869 & 2.0 & 498 & 398 & 37 \\
\bottomrule
\end{tabularx}
\begin{tablenotes}
    \footnotesize
    \item[a] The number of all SDSS galaxies located around each cluster (within $5R_{200}$ and $\Delta z\pm0.03$).
    \item[b] The number of galaxies with SFR and $M_\star$ measurements.
    \item[c] The number of galaxies with SFR and $M_\star$ measurements, and located within $R_{200}$ and $\Delta z\pm0.01$.
\end{tablenotes}
\end{threeparttable}
\end{table*}

\begin{figure}
    \centering
    \includegraphics[width=0.99\linewidth]{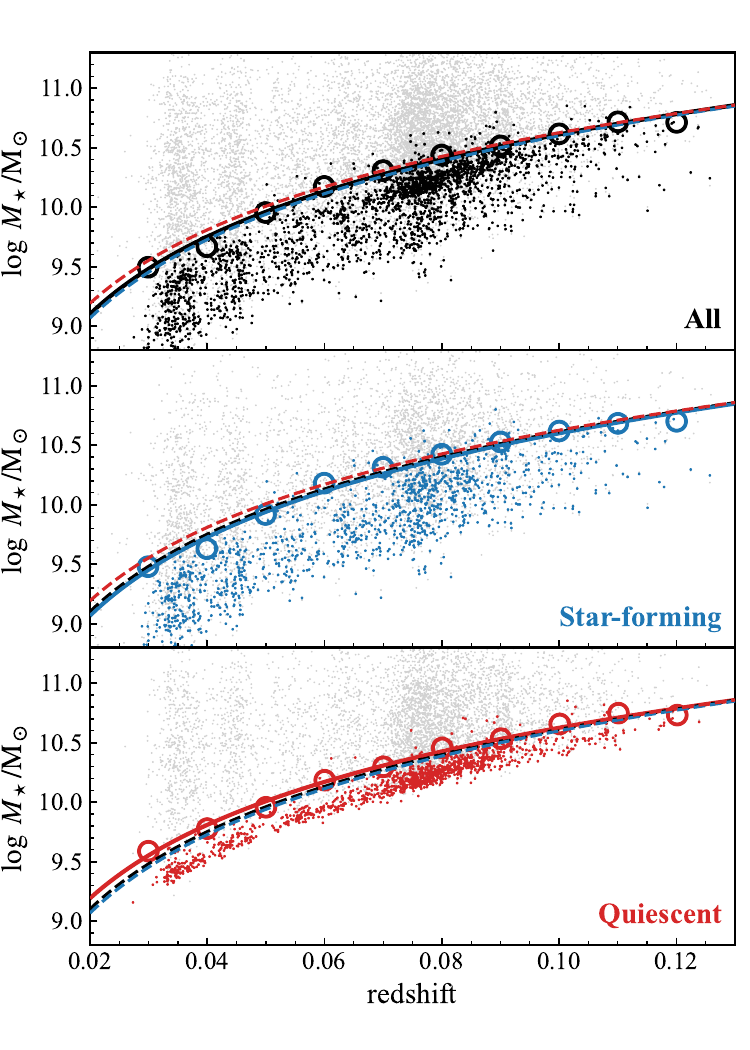}
    \caption{Stellar mass as a function of redshift. The top panel shows the observed stellar masses (gray dots) and the corresponding limiting masses (black dots) of all galaxies. The middle and bottom panels show the same relations for star-forming (blue) and quiescent (red) galaxies, respectively. 
    Open circles mark the 95th percentiles of the limiting-mass distributions in redshift bins of width $\Delta z=0.01$.
    Solid curves show the best-fit completeness relations for the corresponding subsamples, while dashed curves show the fits for the other subsamples for comparison.
    We adopt the quiescent-galaxy fit (red curve) as the stellar-mass completeness limit for the full sample.}
    \label{fig:Mass_comp}
\end{figure}

\subsection{Cluster sample} \label{sec:cls_smpl}

The cluster sample is selected from the Meta-Catalog of X-Ray Detected Clusters of Galaxies (MCXC, \citealt{2011A&A...534A.109P}), which is compiled from seven ROSAT All Sky Survey-based and five serendipitous cluster catalogs, and provides essential clusters properties, including redshift, coordinates, standardized $0.1-2.4$ keV band luminosity $L_{X,500}$, and $R_{500}$ (the radius within which the mean enclosed density is 500 times the critical density of the universe at the cluster redshift).
We convert $R_{500}$ to $R_{200}$ using $R_{200}=R_{500}/0.65$, assuming an NFW profile with a concentration parameter of $c=4$ \citep{Reiprich_2013}.
While assuming a fixed concentration ($c=4$) formally introduces some uncertainty into the derived $R_{200}$, the practical impact is negligible. Observationally, the concentration parameter exhibits a natural scatter (typically $c\sim2-8$) and is also affected by cluster triaxiality \citep{Merten2015}. However, the spatial conversion factor between overdensities is insensitive to the exact concentration. Specifically, the ratio $R_{500}/R_{200}$ varies only from approximately 0.61 (for $c=2$) to 0.68 (for $c=8$). This translates to a minimal systematic variance of $\lesssim5\%$ in $R_{200}$. Therefore, adopting a fixed $c=4$ is highly robust for our phase-space analysis.

We select clusters within a narrow redshift range of $0.03\le z\le0.1$ to minimize the impact of redshift evolution over the sample.
We applied an X-ray luminosity threshold of $L_{\rm X,500}>10^{44}\rm erg/s$ to focus on massive clusters with well-established X-ray cores.

This initial selection yields 140 clusters, and all of them are more massive than $10^{14}\,\rm M_\odot$.

\subsection{Galaxy sample} \label{sec:gal_smpl}

The galaxy sample is drawn from the Sloan Digital Sky Survey Eighteenth Data Release \citep[SDSS-DR18,][]{SDSS_DR18}, the first release for SDSS-V \citep{SDSS_V}. 
We first select galaxies projected within a $5R_{200}$ radius around the cluster centers and with spectroscopic redshifts satisfying $|z - z_{\rm cl}|\le0.03$, where $z_{\rm cl}$ is the cluster redshift.
These criteria provide a sufficient volume to identify surrounding structures and include field galaxies that are not influenced by the cluster environment.
The cluster center coordinates and redshifts are taken from MCXC.

The star formation rates (SFRs) and stellar masses ($M_\star$) of galaxies are taken from the Second Version of GALEX-SDSS-WISE Legacy Catalog \citep[GSWLC-2,][]{Salim_2016, Salim_2018}, which provides physical properties for $\sim700,000$ galaxies at $z<0.3$ derived from SED fitting to UV, optical, and mid-IR photometry.

To ensure sufficient galaxies for robust analysis, we require at least 30 galaxies within $R_{200}$ and within $|z - z_{\rm cl}| \le 0.01$ for each cluster.
We denote this number as $N_{\rm core}$ in Table~\ref{tab:cls_cat}, which also lists the total number of galaxies ($N_{\rm tot}$) and the number of galaxies with both SFR and $M_\star$ measurements ($N_{\rm data}$).
Note that the $N_{\rm core}$ criterion is used solely as a preliminary quality-control step to ensure sufficient data density for each cluster. The actual physical membership classification is strictly determined later in Section \ref{sec:structures}.

In addition, we manually excluded obvious major mergers, including Abell 1650, Abell 2029, and Abell 2033, by visually inspecting their X-ray morphologies.
The final sample contains 25 clusters and 12,498 unique galaxies, of which 11,122 have SFR and $M_\star$ measurements.

\subsection{Stellar mass completeness limit}

The SDSS spectroscopic galaxy sample is magnitude-limited at $r=17.77$ \citep{Strauss2002}. 
However, the stellar-mass completeness limit depends on both redshift and the mass-to-light ratio ($M/L$).
We estimate the mass completeness limit using the empirical method of \citet{Pozzetti2010}.

For each galaxy, we define the limiting stellar mass as the mass it would have if its apparent magnitude were equal to the survey limit, $r_{\rm lim}=17.77$: 
\begin{equation}
\log M_{\rm lim} = \log M_{\rm real}+0.4 (r_{\rm real}-r_{\rm lim})    
\end{equation}
where $M_{\rm real}$ and $r_{\rm real}$ are the observed stellar mass and $r$-band apparent magnitude, respectively.
The distribution of $M_{\rm lim}$ therefore reflects the distribution of galaxy $M/L$.
The real and limiting stellar masses are shown as gray and black dots in the upper panel of Figure~\ref{fig:Mass_comp}.

We divide the sample into redshift bins with a width of 0.01.
In each bin, we compute the 95th percentile of $M_{\rm lim}$. These values are shown as open circles and are adopted as the 95\% completeness limits.
We fit these limits with a logarithmic function of redshift.
We also perform the fit separately for star-forming ($\log_{10}{\rm sSFR}/{\rm yr^{-1}}>-11$; see Section~\ref{sec:SF_Q}) and quiescent galaxies, as shown in the middle and lower panels of Figure~\ref{fig:Mass_comp}
The best-fit completeness curves are shown as dashed lines in all panels and as solid lines in their corresponding panels.
The three curves are quite close to one another, with the quiescent-galaxy curve lying slightly above the other two.

We therefore adopt the quiescent-galaxy fit (red curve) as the stellar-mass completeness limit for the full sample:
\begin{equation}
    \log_{10}M_{\rm limit}(z)/{\rm M_\odot} = 2.05\cdot\log_{10} z + 12.68
\end{equation}
Galaxies above the red curve are defined as the mass-complete sample.
Because few galaxies in the mass-complete sample have masses lower than $10^{9.5}M_{\odot}$, we adopt $10^{9.5}M_{\odot}$ as a mass cut to ensure robust statistics.
Finally, the mass-complete sample contains 7092 galaxies (2654 star-forming and 4438 quiescent), which are used in the analysis in Section~\ref{sec:results}.

\subsection{Classification of star-forming and quiescent galaxies} \label{sec:SF_Q}

\begin{figure*}
    \centering
    \includegraphics[width=0.99\linewidth]{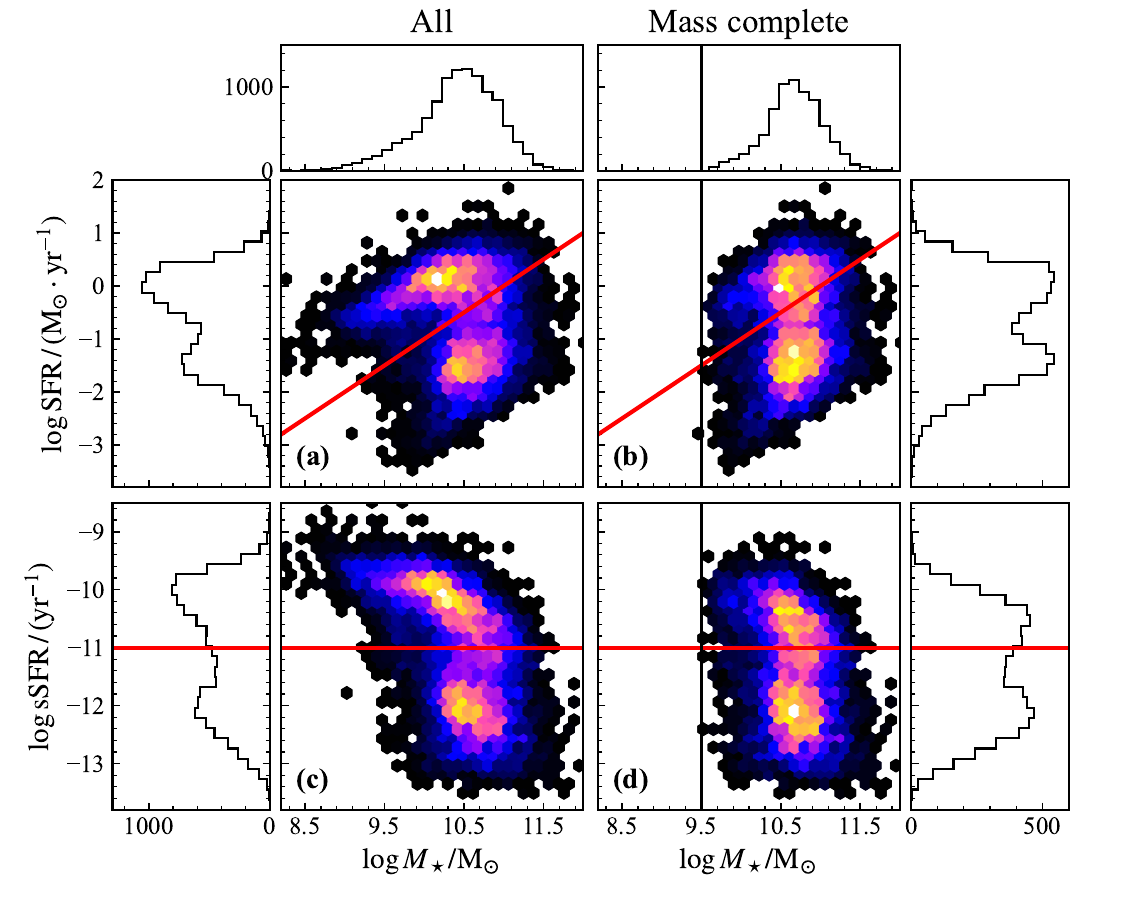}
    \caption{Star formation activity as a function of stellar mass for the full sample (left column) and the mass-complete sample (right column). Panels (a) and (b) show $\log{\rm SFR}$ versus $\log(M_\star/M_\odot)$, while panels (c) and (d) show $\log{\rm sSFR}$ versus $\log M_\star/M_\odot$. The red lines indicate the adopted threshold of $\log_{10}{\rm sSFR}/{\rm yr^{-1}}=-11$ for separating star-forming and quiescent galaxies. The solid black lines mark the stellar-mass cut of $\log M_\star/M_\odot=9.5$. Marginal histograms show the corresponding one-dimensional distributions.}
    \label{fig:SFR_M}
\end{figure*}

Galaxies exhibit a bimodal distribution in the $\log{\rm SFR}$ vs. $\log M_\star$ plane.
As shown in panel (a) of Figure~\ref{fig:SFR_M}, galaxies with relatively high SFRs follow the tight and nearly linear star-forming main sequence \citep[SFMS, e.g.,][]{Salim_2007}. A distinct population of quiescent galaxies with significantly lower SFRs is also evident, particularly at the high-mass end.
The distribution of specific star formation rate (sSFR), defined as $\log{\rm sSFR}=\log{\rm SFR}-\log M_\star$, is shown as a function of stellar mass in panel (c) of Figure~\ref{fig:SFR_M}.
To separate star-forming and quiescent galaxies, we adopt a commonly used threshold of $\log_{10}{\rm sSFR}/{\rm yr^{-1}} = -11$ \citep[e.g.,][]{Brinchmann2004, 2008Franx, Fontanot2009, McGee2011, 2013Ilbert, Sherman2020, Donnari_2021}, as indicated by the red solid lines. 

The SFR and sSFR distributions of the mass-complete sample are shown in panel (b) and (d) of Figure~\ref{fig:SFR_M}, respectively.
The mass-complete selection preferentially removes low-mass star-forming galaxies since they generally have higher $M/L$ than quiescent galaxies.
Note that although a slightly lower threshold (e.g., $\log_{10}{\rm sSFR}/{\rm yr^{-1}}\approx -11.2$) might be closer to the saddle point of the sSFR distribution, according to panel (d) and the corresponding histogram, we adopt the more commonly used $\log_{10}{\rm sSFR}/{\rm yr^{-1}} = -11$ threshold to facilitate direct comparisons with previous studies.

\section{Methods}
\label{sec:methods}

\subsection{Tracing infall process in the $R$--$V$ diagram}
\label{sec:infall}

\begin{figure}
    \centering
    \includegraphics[width=0.9\linewidth]{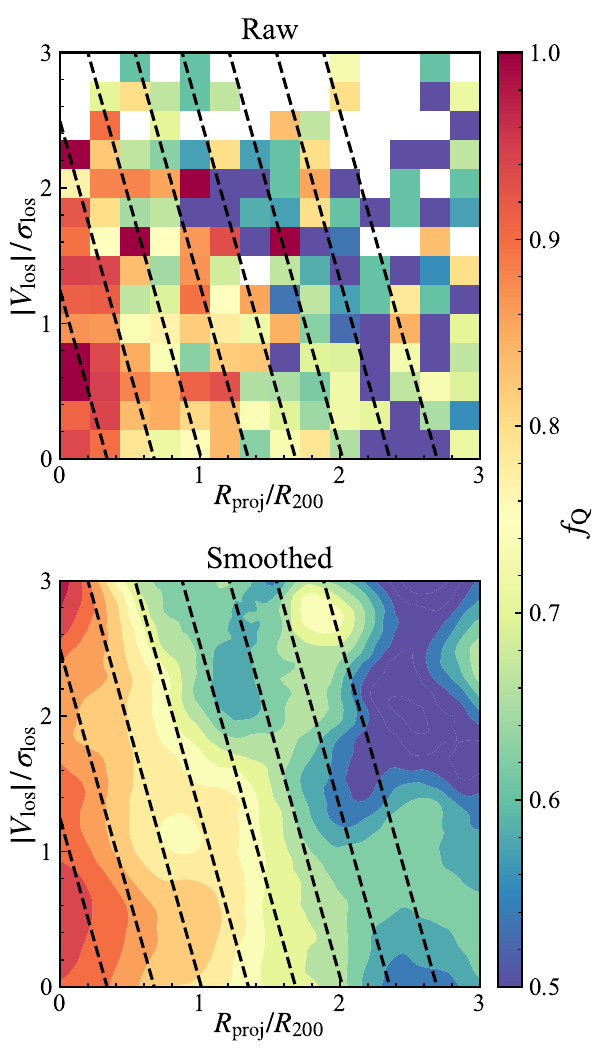}
    \caption{Quiescent fraction $f_{\rm Q}$ in the $R$--$V$ diagram. 
    Top: the raw, binned distribution in the $R$--$V$ diagram. Bins containing fewer than five galaxies are left blank (white). 
    Bottom: the same distribution after smoothing.
    Dashed lines illustrate the parallel lines \citep[slope $k=-3.7$,][]{Dou25} used to define the infall proxy $d_{\rm R}$.
    }
    \label{fig:fQ_PPS}
\end{figure}

To explore galaxy quenching process as a function of infall stage, we employ a statistical infall proxy provided by \citet{Dou25}, which uses a series of parallel lines to trace the infall process in the $R$--$V$ diagram. 
Using the TNG300 simulation, \citet{Dou25} demonstrated that this linear proxy provides a more accurate correspondence with infall time than previous methods. 
Examples of these parallel lines are shown as dashed lines in Figure~\ref{fig:fQ_PPS}.

For each galaxy, there is a corresponding line passing through its location.
\citet{Dou25} originally used the perpendicular distance from the origin to the line, $d_{\rm linear} = (|V| + kR)/\sqrt{1+k^2}$, to quantify the infall process.
Here we instead use the $x$-intercept of the line:
\begin{equation}
    d_{\rm R} = R - |V|/k
\end{equation}
where $k=-3.7$ is the slope of the line, $R = R_{\rm proj}/R_{200}$ is the projected clustercentric radius normalized by $R_{200}$, and $V = |\Delta V_{\rm los}|/\sigma$ is the line-of-sight velocity offset from the cluster center normalized by the cluster velocity dispersion. The velocity dispersion is estimated using the caustic method \citep{Diaferio1997, Diaferio1999, Serra2011}.
With this definition, smaller $d_{\rm R}$ corresponds to earlier infall, and $d_{\rm R}$ is convenient for comparing trends along infall stage with those as a function of the projected radius.

The distribution of the quiescent fraction ($f_{\rm Q}$) in the $R$--$V$ diagram is shown in Figure~\ref{fig:fQ_PPS}. The top panel presents the raw binned map, and the bottom panel shows the same distribution smoothed using an Epanechnikov kernel with a size of 0.5. 
The parallel lines from \citet{Dou25} broadly follow the contours of $f_{\rm Q}$, supporting the use of $d_{\rm R}$ as a practical proxy for tracing evolutionary trends during cluster infall.

\subsection{Structure identification}
\label{sec:structures}

\begin{figure*}
    \centering
    \includegraphics[width=0.9\linewidth]{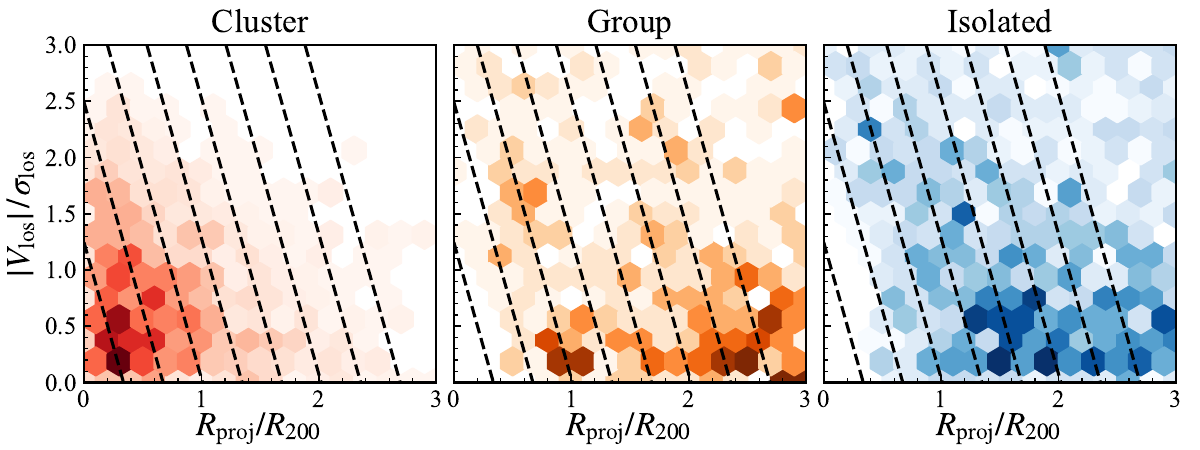}
    \caption{ The distribution in the $R$--$V$ diagram for cluster galaxies (red), group galaxies (orange), and isolated galaxies (blue), respectively, with colors indicating the number densities. The dashed lines illustrate the parallel lines of \citet{Dou25}.}
    \label{fig:PPS}
\end{figure*}

Cluster galaxies may have experienced multiple environments before accretion, implying diverse infall pathways. 
To investigate the impact of such pathways, we classify galaxies according to their local environments.

We adopt the Blooming Tree (BT) algorithm \citep{Yu2018, Yu2025} to identify structures around each cluster.
This algorithm arranges all the galaxies in the field of view into a dendrogram, based on the pairwise projected binding energy, which is estimated from the location of the galaxies on the sky and their redshift \citep{Diaferio1999, Yu2015, Yu2016}. 
The dendrogram is subsequently trimmed into distinct structures using a density contrast parameter $\Delta \eta$, which quantifies the degree to which a structure is overdense relative to the background. 
The $\eta$ parameter itself reflects the compactness of a structure, taking into account its size, number of member galaxies, and line-of-sight velocity dispersion.
Larger density contrast $\Delta \eta$ selects increasingly dense and gravitationally bound structures.
The specific choice of $\Delta \eta$ also depends on the sampling density. For SDSS, the BT algorithm identifies superclusters with $\Delta \eta=1$, clusters with $\Delta \eta=20$, and dense core of clusters with $\Delta \eta=50$ \citep{Yu2025}.

We adopt $\Delta\eta=20$ to identify all structures in the field of view. Then galaxies belonging to the most central structure of each cluster are classified as core members and are hereafter referred to as ``cluster galaxies''.
They are fundamentally dominated by the main dark halo of the cluster.
Galaxies residing in other identified structures are defined as ``group galaxies''. 
These systems are dominated by local group-scale halos, including both infalling groups surrounding the clusters and substructures within the clusters.
Galaxies not associated with any identified structure are classified as ``isolated galaxies'', including galaxies that are close to the cluster but infalling individually, and real field galaxies that have no physical association with the cluster.

Applying this classification to the 7092 mass-complete galaxies yields 1405 cluster galaxies, 1737 group galaxies, and 3950 isolated galaxies.
All 12,498 galaxies with spectroscopic redshifts are used for structure identification, whereas only the mass-complete sample is used in the subsequent analysis.

The distributions of the three populations in the $R$--$V$ diagram are presented in Figure~\ref{fig:PPS}.
Cluster galaxies preferentially occupy the virialized region, whereas group and isolated galaxies are more common at larger radii and larger velocity.
However, a small number of cluster galaxies extend to large radii but have small velocities. We attribute this to structure misclassification and discuss it further in Section~\ref{sec:interloper}.

\section{Results} \label{sec:results}

In the subsequent analysis, the quiescent fraction ($f_{\rm Q}$) within a specific parameter bin is simply computed as the raw ratio of number of quiescent galaxies to the total number of galaxies.
Uncertainties in $f_{\rm Q}$ are computed using the Wilson score interval \citep{Wilson1927, astropy} for a binomial proportion, and the corresponding $1\sigma$ (68\%) confidence limits are shown as shaded regions in each figure.

\subsection{Quenching of all galaxies}

\begin{figure}[ht]
    \centering
    \includegraphics[width=0.99\linewidth]{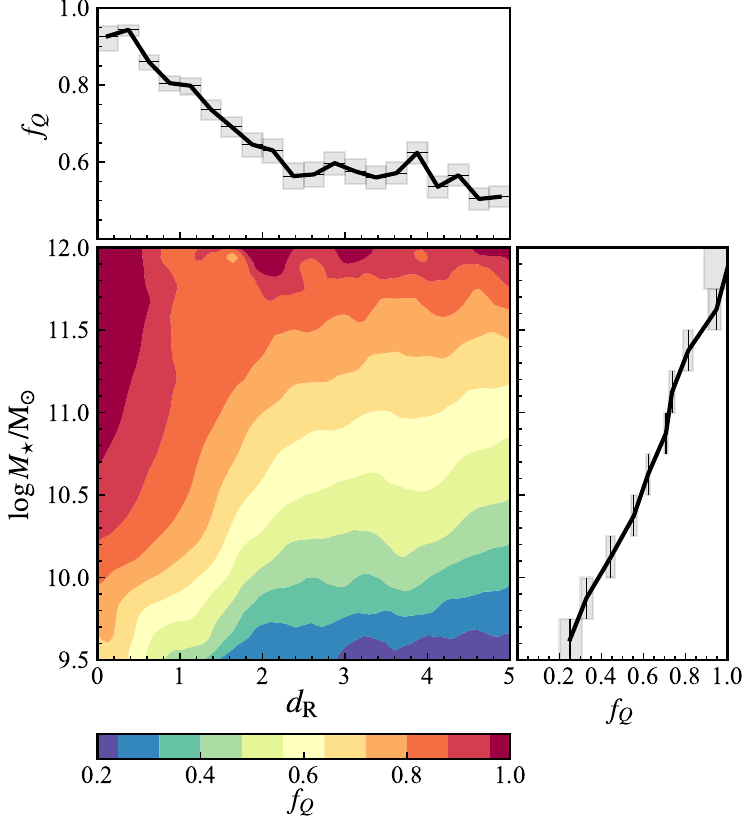}
    \caption{Quiescent fraction $f_{\rm Q}$ as a function of stellar mass and the infall proxy $d_{\rm R}$ for the full galaxy sample. 
    The bottom-left panel shows $f_{\rm Q}$ in the $\log M_\star/M_\odot-d_{\rm R}$ plane with colors indicating $f_{\rm Q}$. 
    The top and right panels show the corresponding one-dimensional trends of $f_{\rm Q}$ as a function of $d_{\rm R}$ and $\log M_\star/M_\odot$, respectively. 
    The bin sizes $\Delta d_{\rm R}$ and $\Delta \log M_\star/M_\odot$ are both 0.25.
    The $f_{\rm Q}$ values in each bin are indicated by the solid horizontal lines and the shaded squares show the 68\% Wilson binomial confidence intervals on $f_{\rm Q}$.
    }
    \label{fig:f_Q}
\end{figure}

Since both internal and external processes can quench galaxies, we examine both by plotting the quiescent fraction $f_{\rm Q}$ as a function of stellar mass and the infall proxy $d_{\rm R}$ in the bottom left panel of Figure~\ref{fig:f_Q}, with the corresponding one-dimensional trends shown in the top and right panels, and uncertainties indicated by the gray squares.
As expected, $f_{\rm Q}$ increases with stellar mass: galaxies are more likely to be quiescent at higher $M_\star$, consistent with stronger internal feedback in more massive systems.
Similar results have been reported in many previous works \citep[e.g.,][]{Kauffmann2004,Peng_2010,Taylor_2023,Shi_2024}, and the most massive galaxies ($\log M_\star>10^{11.5}\rm M_\odot$) are nearly all quiescent even in the field.

The quiescent fraction shows a two-stage behavior along the infall process. 
At the early stage of infall ($d_{\rm R}\gtrsim2.5$), the quiescent fraction keeps nearly constant.
Once galaxies reach $d_{\rm R}\approx2.5$, $f_{\rm Q}$ increases steadily up to nearly 100\% toward the cluster center.
This behavior suggests a characteristic boundary of clusters, inside which environmental effects become increasingly important for quenching.

\subsection{Quenching in three different local environments}

\begin{figure*}[ht]
    \centering
    \includegraphics[width=0.45\linewidth]{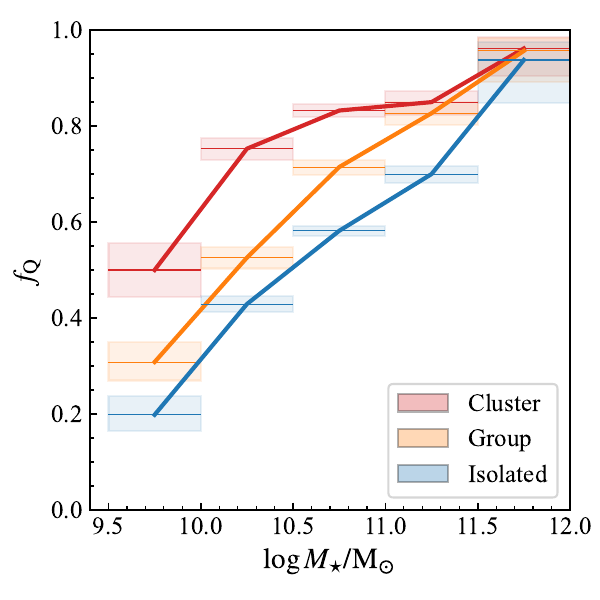}
    \includegraphics[width=0.45\linewidth]{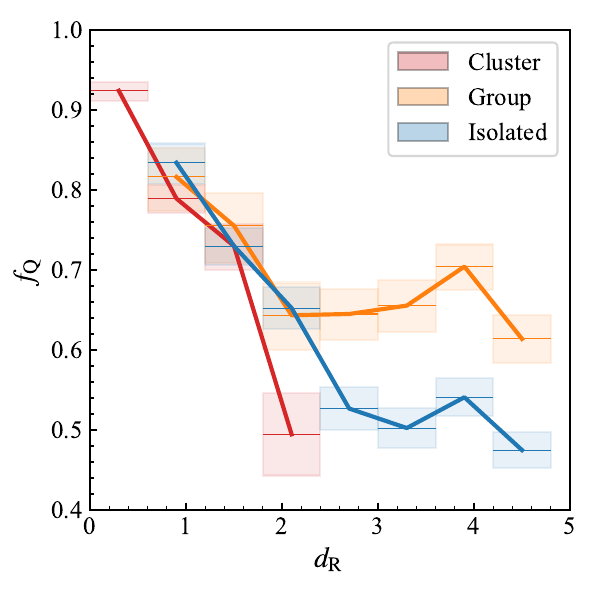}
    \caption{Quiescent fraction $f_{\rm Q}$ for galaxies in three local environments as a function of stellar mass (left) and the infall proxy $d_{\rm R}$ (right). 
    The bin sizes are $\Delta \log M_\star/M_\odot=0.5$ and $\Delta d_{\rm R}=0.6$, respectively.
    Cluster, group, and isolated galaxies are shown by the red, orange, and blue curves, respectively. 
    Shaded squares indicate the uncertainties in $f_{\rm Q}$.}
    \label{fig:f_Q_3sam}
\end{figure*}

Here we examine quenching in cluster, group, and isolated galaxies separately.
The overall quiescent fractions of cluster, group, and isolated galaxies are 80\%, 65\% and 55\%, respectively, consistent with more efficient quenching in denser environments.
The quiescent fraction as a function of stellar mass and the infall proxy $d_{\rm R}$ are shown in the left and right panels of Figure~\ref{fig:f_Q_3sam}, with cluster, group, and isolated galaxies shown as red, orange, and blue curves, respectively. 

In the $f_{\rm Q}$--$M_\star$ relation, the group galaxies exhibit a systematically higher quiescent fraction (by roughly 10\%) than the isolated galaxies across the entire mass range, and the two curves show remarkably similar slopes. This nearly constant offset suggests that the additional quenching driven by the group-scale environment has a broadly comparable efficiency across the stellar masses probed here.
In contrast, cluster galaxies show a different mass dependence.
They have quiescent fractions similar to those of group galaxies at the high-mass end ($\log M_\star/\rm M_\odot>11$), but display significantly higher quiescent fractions at the low-mass end, where $f_{\rm Q}$ reaches approximately $50\%$ at $\log M_\star/\rm M_\odot=9.75$, compared to $f_{\rm Q}\approx 30\%$ for group galaxies.
Such a mass dependence is consistent with more efficient cluster-driven quenching (e.g., ram pressure stripping) for lower-mass galaxies, which have shallow potentials. 
At the high-mass end, galaxies are less sensitive to environment and are often quenched primarily by internal processes.

Turning to the $f_{\rm Q}$--$d_{\rm R}$ relation, all three samples broadly follow the overall trend in Figure~\ref{fig:f_Q}, showing a outer plateau followed by a rise toward smaller $d_{\rm R}$.
The cluster sample is an exception, with $f_{\rm Q}$ increasing steadily from $\sim 50\%$ at $d_{\rm R}\approx 2$ to $>90\%$ toward the center. Because cluster galaxies are mainly concentrated at $d_{\rm R}<2$, the number of cluster galaxies at larger $d_{\rm R}$ is small and the outer trend is therefore less well constrained.
A notable feature is the difference between group and isolated galaxies.
Both show an outer plateau, but at different levels (approximately 65\% and 50\%, respectively), suggesting the ``pre-processing'' in group-scale halos prior to cluster infall.

More importantly, the transition points of group and isolated galaxies differ: $f_{\rm Q}$ for isolated galaxies begins to rise at $d_{\rm R}\approx 2.5$, whereas the group sample remains flat until $d_{\rm R}\lesssim 2$, indicating a delayed quenching of group galaxies.
Additional evidence of this delay lies in the overall trends: although isolated galaxies exhibit a quiescent fraction $\sim 15\%$ lower than group galaxies at $d_{\rm R}>3$, they catch up and converge with group galaxies within $d_{\rm R}=2$, suggesting a higher quenching efficiency of isolated galaxies than group galaxies.

At $d_{\rm R}<2$, all three populations exhibit indistinguishable quiescent fractions within uncertainty, indicating that the overwhelming cluster environment completely dominates galaxy quenching and washes out the initial differences imprinted by the local group-scale environments.

\subsection{The influence of mass on the quenching process}
\label{sec:mass_bin}

\begin{figure*}
    \centering
    \includegraphics[width=0.99\linewidth]{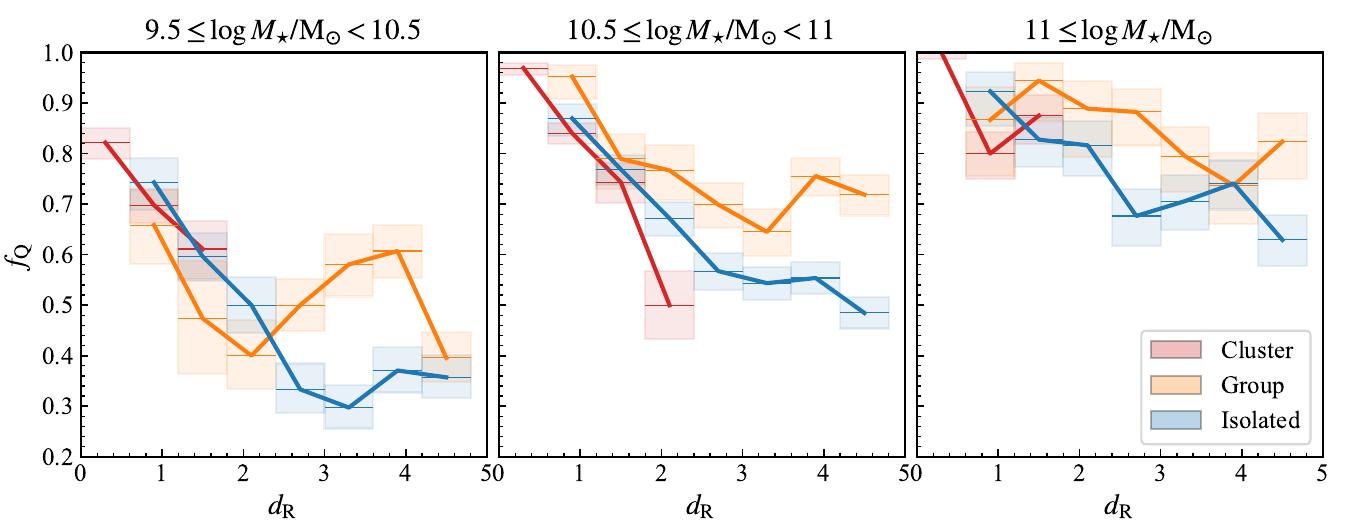}
    \caption{Quiescent fraction $f_{\rm Q}$ as a function of the infall proxy $d_{\rm R}$ for galaxies across different local environments and different stellar-mass bins. The bin size is $\Delta d_{\rm R}=0.6$.}
    \label{fig:massbin}
\end{figure*}

To further investigate how stellar mass and environment jointly influence quenching during infall, we subdivide each of the three environmental samples into three stellar-mass bins: low-mass galaxies with $9.5\le\log M_\star/\rm M_\odot<10.5$, intermediate-mass galaxies with $10.5\le\log M_\star/\rm M_\odot<11$, and high-mass galaxies with $\log M_\star/\rm M_\odot\ge11$.
The resulting $f_{\rm Q}$--$d_{\rm R}$ relations are shown in Figure~\ref{fig:massbin}.

Overall, more massive galaxies exhibit systematically higher $f_{\rm Q}$ across all environments, consistent with the trends shown in Figures~\ref{fig:f_Q} and~\ref{fig:f_Q_3sam}.
At fixed mass, the signatures of ``pre-processing'' and the delayed quenching of group galaxies observed in Section~\ref{sec:results} remain visible, although their amplitudes vary with stellar mass.

The high-mass galaxies show mutually consistent trends within uncertainties, suggesting that internal rather than environmental quenching dominates at $\log (M_\star/M_\odot)\ge 11$.
In contrast, for lower-mass galaxies, both ``pre-processing'' and the delayed quenching of group galaxies are significantly more pronounced.
In the low-mass sample, the $f_{\rm Q}$ offset between group and isolated galaxies reaches $>20\%$ at $d_{\rm R}>3$ but fully converges at $d_{\rm R}<2.5$. The intermediate-mass galaxies exhibit transitional behavior that bridges the low- and high-mass cases.

\section{Discussion}
\label{sec:discussion}

\subsection{Physical meaning of $d_{\rm R}$
} \label{sec:discuss_dR}

Throughout this paper, we use $d_{\rm R}$ as a proxy to trace the infall process. However, $d_{\rm R}$ does not provide a one-to-one mapping to the true infall time.
\citet{Dou25} demonstrated that infall times inferred from the $R$--$V$ diagram exhibit substantial intrinsic dispersion owing to orbital overlap in the $R$--$V$ diagram and projection effects.
As a result, although a given $d_{\rm R}$ value is associated with a characteristic infall stage, it actually corresponds to a broad distribution of infall times.
The evolutionary trends reported as a function of $d_{\rm R}$ should therefore be interpreted as population-averaged behavior, tracing the gradual transition from the unperturbed cosmic field to the dense cluster environment, rather than as the evolution of individual galaxies along a single, well-defined timeline.

\subsection{The ``delay-then-rapid'' scenario}

As shown in Section~\ref{sec:results}, the variation in the quiescent fraction along $d_{\rm R}$ exhibits two phenomenological stages: a plateau at the beginning of infall and a steady quenching after a transition point.
This behavior is broadly consistent with the well known ``delay-then-rapid'' quenching scenario \citep{Wetzel_2013, Haines_2013_slowquench}.
In this framework, galaxies are not quenched immediately upon their first infall into a host halo. Instead, they maintain roughly constant star formation for $\sim2-4$~Gyr, a phase during which ``starvation'' (i.e., the shutoff of gas replenishment) is thought to dominate. After this delay phase, galaxies undergo rapid quenching on a timescale of $\lesssim1$ Gyr, as more efficient mechanisms such as ram pressure stripping become important.

Although the original ``delay-then-rapid'' scenario was formulated with respect to a galaxy’s first infall into any host halo (often during the group stage), rather than the infall into the current (final) cluster halo considered here, our results can be interpreted within a similar physical framework.
At large $d_{\rm R}$ ($\gtrsim 2.5$), corresponding to the early stage of cluster infall, the ICM is relatively diffuse and is more likely to primarily affect the extended gas reservoir (i.e., the CGM) of galaxies, leading to starvation.
As galaxies move to smaller $d_{\rm R}$ ($\lesssim 2.5$), they encounter denser regions where ram pressure stripping becomes increasingly effective, driving $f_{\rm Q}$ upward toward the cluster center.

\subsection{The dual role of the group-scale environments}

It has long been recognized that galaxies in clusters are influenced not only by their current cluster environment but also by the environments they experienced in histories. 
In particular, galaxies residing in group-scale halos often show suppressed star formation relative to field galaxies, even before reaching the densest regions of clusters. This ``pre-processing'' has been extensively studied \citep{Fujita_2004, 2013Vijayaraghavan, Kleiner_2021, Piraino_2024, Lopes24}.

Our analysis reveals an additional, distinct effect associated with group environments. 
As demonstrated in Figure~\ref{fig:f_Q_3sam}, group galaxies exhibit a delayed quenching compared to isolated galaxies.  
This suggests that group galaxies are less rapidly affected than isolated galaxies upon entering the cluster.
This behavior can be phenomenologically described as a ``protection'' effect of the group environments.
Moreover, we find that this ``protection'' effect depends on stellar mass, appearing stronger at lower stellar masses.
This is consistent with the scenario where massive galaxies are primarily quenched by internal processes, whereas low-mass galaxies are more sensitive to environmental mechanisms.

The dual roles of ``pre-processing'' and ``protection'' are fundamentally compatible. 
Relative to the field, the denser group environment enhances quenching, thereby increasing the baseline $f_{\rm Q}$ prior to infall.
Relative to the harsh cluster environment, however, group-scale halos may serve as a buffer that reduces the immediate impact of cluster-related processes on their member galaxies.
In contrast, isolated galaxies are directly exposed to the ICM and the cluster potential immediately upon infall.

Similar effect has been explored in The Three Hundred simulation. \citet{2022Kotecha} found that within the hot dense cluster environment, galaxies closer to intra-cluster filaments exhibit delayed quenching compared to those outside filaments. 
The underlying physical mechanisms are complex.
\citet{2022Kotecha} examined the gas flow and found that streams of cold gas moving coherently with the galaxies can effectively decrease the local ram pressure toward centers of filaments, thereby shielding them from gas stripping.
Furthermore, this gas can be accreted onto galaxies, serving as fuel for sustained star formation.
Given that galaxy groups generally infall into clusters along filaments, the ``protection'' effect observed in our group galaxies may be physically associated with filaments.
Moreover, \citet{2022Kotecha} also noted that within clusters, more massive halos retain a higher fraction of cold gas, whereas smaller halos are more susceptible to the cluster environment. Consequently, the deeper potential wells of the group-scale halos may also assist in retaining cold gas compared to those isolated galaxies.

\subsection{Independent evidence from stellar ages}

\begin{figure*}
    \centering
    \includegraphics[width=0.9\linewidth]{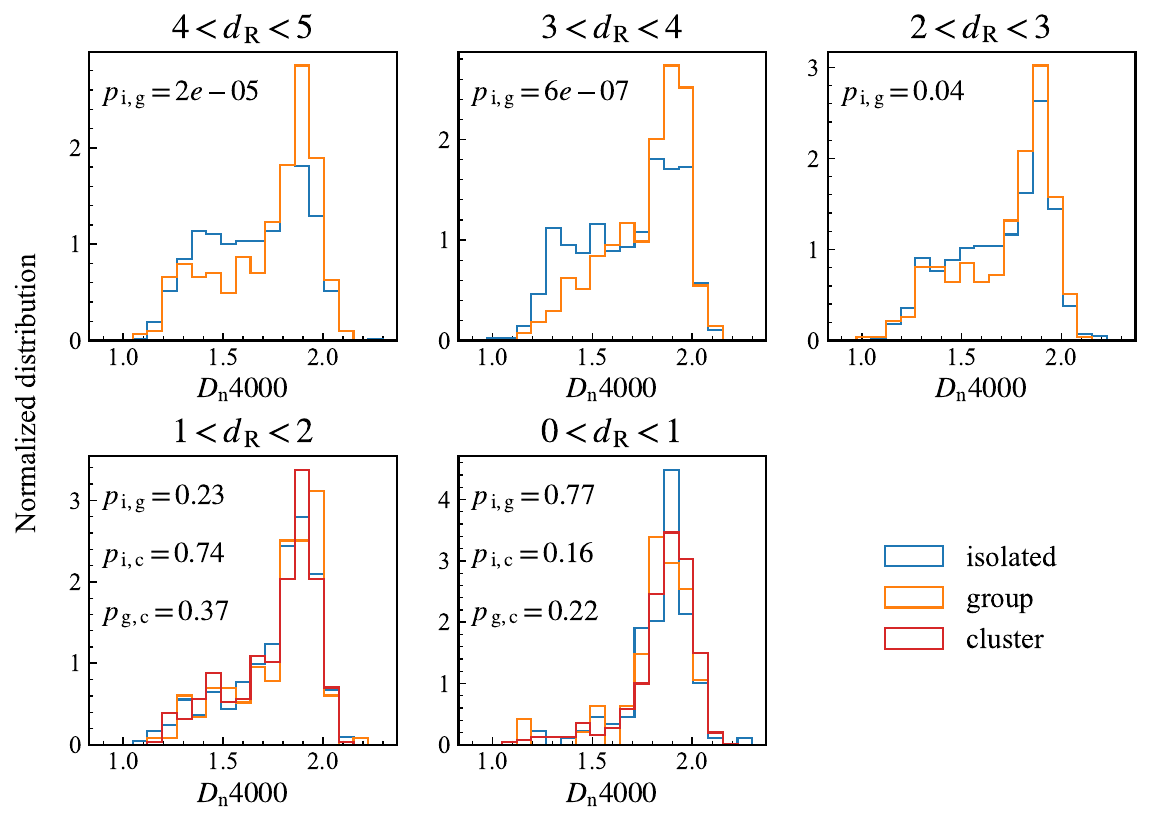}
    \caption{Distributions of the $D_{\rm n}4000$ index for cluster (red), group (orange), and isolated (blue) galaxies across different $d_{\rm R}$ intervals. The $p$-values derived from two-sample Kolmogorov-Smirnov (KS) tests comparing these populations are provided in the upper left corner of each panel. The subscripts 'i', 'g', and 'c' denote the isolated, group, and cluster galaxies, respectively.}
    \label{fig:Dn4000}
\end{figure*}

Quenching mechanisms of ``pre-processing'', such as starvation, typically require a substantial timescale to suppress the star-formation rate sufficiently to visibly alter the quiescent fraction, $f_Q$. Therefore, the elevated $f_Q$ observed in group galaxies at early infall stages ($d_{\rm R} > 3$) strongly implies that these galaxies have already experienced a prolonged ``pre-processing'' history within their local group halos prior to the cluster accretion. To provide independent spectroscopic evidence for this long-term evolutionary divergence, we examine the distribution of the $D_{\rm n}4000$ index across different $d_{\rm R}$ as shown in Figure~\ref{fig:Dn4000}. As a robust proxy for the mean stellar age, the $D_{\rm n}4000$ index integrates the star-formation history over timescales of $\sim1-2$~Gyr \citep[e.g.,][]{Kauffmann2004}, allowing us to track the cumulative effect of environments. We also performed Kolmogorov-Smirnov (KS) tests to quantify the statistical significance of the differences between these distributions, with the corresponding $p$-values indicated in the figure.

At the early infall stages ($d_{\rm R} > 3$), group galaxies exhibit a systematic shift toward larger $D_{\rm n}4000$ values compared to isolated galaxies, with extremely small $p$-values indicating significant difference between these two populations. This contrast suggests older stellar populations of group galaxies and provides independent spectroscopic evidence for ``pre-processing''.
Furthermore, as galaxies move toward the cluster center (smaller $d_{\rm R}$), the $D_{\rm n}4000$ distributions of all three populations gradually converge. Within $d_{\rm R} < 2$, the distributions become virtually indistinguishable ($p$-values $>0.05$), with the majority of galaxies reaching high $D_{\rm n}4000$ values. This evolutionary convergence mirrors the behavior seen in the $f_{\rm Q}$--$d_{\rm R}$ relation (Figure~\ref{fig:f_Q_3sam}) and reinforces the conclusion that the overwhelming cluster-driven mechanisms (e.g., ram pressure stripping) ultimately dominate the quenching process, effectively washing out the initial differences imprinted by their pre-cluster accretion histories.

\subsection{Contamination from interlopers and backsplash galaxies} \label{sec:interloper}

In the $R$--$V$ diagram, contamination from interlopers is physically unavoidable because of projection effects, especially at $R\gtrsim2$ \citep{Oman_2016,Dou25}. Interlopers are foreground or background galaxies that have no physical association with the target cluster. However, our methodology effectively mitigates this contamination. Because our algorithm identifies groups and clusters based on robust binding energy and local overdensities, dynamically unbound interlopers typically fail to meet the membership criteria for these dense structures. Consequently, the vast majority of them are assigned to the isolated sample, preserving the purity of the group and core cluster populations.
The only notable exception occurs at $d_{\rm R} \approx 2$ for cluster galaxies. At this large clustercentric distance, true cluster members become intrinsically sparse and this specific radial bin in the cluster sample may be primarily dominated by interlopers, leading to the low $f_{\rm Q}$ shown in Figure~\ref{fig:f_Q}. 

Furthermore, it is crucial to recognize the diluting impact of any residual contamination. The quiescent fraction of group galaxies will be lower than the genuine value if they are contaminated by interlopers, thus reducing the offset between group and isolated galaxies we have observed. Therefore, the conclusions about ``pre-processing'' and ``protection'' can be safely regarded as a conservative lower limit and remain qualitatively robust.

Backsplash galaxies are those which have passed through the cluster core for at least once and rebounded to the outskirts  \citep[typically out to the splashback radius of $R_{\rm sp}\lesssim 2R_{200}$][]{More2015, More2016}. 
Even if they infall as part of a group at the beginning, their structures are largely disrupted by severe tidal stripping during their pericentric passages.
Thus most of them are classified as cluster galaxies (the core structure) by the BT algorithm because their are close to the center and without local binding. While some residuals bouncing very far are classified as isolated galaxies, leading to an elevated quiescent fraction than its true value.

\section{Summary}
\label{sec:summary}

In this paper, we investigate galaxy quenching during cluster infall and assess how local group-scale environments modulate this process.
We utilized a sample of 25 low-redshift, X-ray luminous massive clusters selected from MCXC, combined with a mass-complete galaxy sample covering $5R_{200}$ around the clusters.

To trace infall stage, we adopted the infall proxy $d_{\rm R}$, defined by a series of parallel lines in the $R$--$V$ diagram. 
We emphasize that $d_{\rm R}$ serves as a phenomenological metric tracing the macroscopic infall process rather than an accurate clock.
To distinguish local environments, we applied the Blooming Tree algorithm to identify substructures based on the projected binding energy and classified galaxies into three populations: cluster galaxies (members of the core structure), group galaxies (members of other structures), and isolated galaxies (not assigned to any structure). 
This physical binding requirement effectively mitigates contamination from purely projected interlopers.

The quiescent fraction generally exhibits a clear two-stage behavior as a function of $d_{\rm R}$. It remains approximately constant at $d_{\rm R}\gtrsim 2.5$ and then rises steadily toward smaller $d_{\rm R}$, approaching unity near the cluster centers. 
Phenomenologically, this is consistent with the ``delay-then-rapid'' scenario in which relatively gentle processes operate during the early infall stages, while more efficient environmental mechanisms become increasingly dominant at smaller $d_{\rm R}$.

Furthermore, our results reveal that group-scale environments play a dual role in quenching.
At large $d_{\rm R}$, group galaxies show a higher quiescent fraction than isolated galaxies, consistent with ``pre-processing'' in group-scale halos prior to cluster infall. In contrast, the rise of $f_{\rm Q}$ toward smaller $d_{\rm R}$ occurs later for group galaxies compared to isolated galaxies. We refer to this delay as a ``protection'' signature associated with group-scale environments, which may be attributed to the presence of cold gas in filaments.
Finally, at $d_{\rm R}<2$, the quiescent fractions of all three populations converge, indicating that the overwhelming cluster environment ultimately dominates galaxy quenching.

The new generation of wide-field surveys such as DESI \citep{Levi_2013, DESI_2016a} and Euclid \citep{Laureijs_2011, Euclid_2024} will deliver significantly larger and deeper spectroscopic and photometric data sets with improved estimates of galaxy physical properties.
These data will enable more stringent tests of how accretion histories and group-scale environments regulate quenching during cluster infall.

\begin{acknowledgements}
This work has been supported by the National Natural Science Foundation of China No. 12573003, the National Key Research and Development Program of China (No. 2023YFC2206704), 
and the China Manned Space Program with grant No. CMS-CSST-2025-A04.
This work was made possible thanks to a number of open-source software packages: AstroPy \citep{astropy}, Matplotib \citep{matplotlib}, NumPy \citep{numpy}, Pandas \citep{pandas} and SciPy \citep{scipy}.
\end{acknowledgements}

\bibliography{references}{}
\bibliographystyle{aa}

\end{CJK*}
\end{document}